\documentclass{birkjour}

\usepackage{amsmath}
\usepackage{amssymb}
\usepackage{amsthm}
\usepackage{graphicx}

\newtheorem{thm}{Theorem}[section]

\newtheorem{prop}[thm]{Proposition}
\theoremstyle{definition}

\theoremstyle{remark}
\newtheorem{rem}[thm]{Remark}

\numberwithin{equation}{section}

\begin{document}

\firstpage{345}

\title[Atom interferometry]{Equivalence principle, quantum mechanics, and atom-interferometric tests}

\author[D.\ Giulini]{Domenico Giulini}

\address{%
Institute for Theoretical Physics\\
Appelstra\ss e 2\\
D-30167 Hannover\\
Germany\\[1ex]
Center of Applied Space Technology and Microgravity\\
Am Fallturm\\
D-28359 Bremen\\
Germany
}
\email{giulini@itp.uni-hannover.de}

\thanks{I thank Felix Finster, Olaf M\"uller, Marc Nardmann, J\"urgen Tolksdorf and Eberhard Zeidler for inviting me to the conference on \emph{Quantum field theory and gravity} (September~27 -- October~2, 2010). This paper is the written-up and updated version of my talk: \emph{`Down-to-Earth issues in atom interferometry}, given on October 1st 2010. I thank Claus L\"ammerzahl and Ernst Rasel for discussions concerning the issues and views presented here, and Marc Nardmann ans Laura Schmidt for carefully
reading the manuscript. Last not least I sincerely thank the QUEST 
cluster for making this work possible.}



\begin{abstract}
That gravitation can be understood as a purely metric phenomenon depends
crucially on the validity of a number of hypotheses which are
summarised by the Einstein Equivalence Principle, the least well
tested part of which being the Universality of Gravitational Redshift.
A recent and currently widely debated proposal (Nature 463 (2010)
926-929) to re-interpret some 10-year old experiments in atom
interferometry would imply, if tenable, substantial reductions on
upper bounds for possible violations of the Universality of
Gravitational Redshift by four orders of magnitude. This interpretation,
however, is problematic and raises various compatibility issues
concerning basic principles of General Relativity and Quantum Mechanics.
I review some relevant aspects of the equivalence principle and its
import into quantum mechanics, and then turn to the problems raised by
the mentioned proposal. I conclude that this proposal is too problematic
to warrant the claims that were launched with it.
\end{abstract}

\maketitle

\section{Introduction}
\label{sec:Introduction}
That gravitation can be understood as a purely metric phenomenon
depends crucially on the validity of a number of hypotheses which
are summarised by the Einstein Equivalence Principle, henceforth
abbreviated by EEP. These assumptions
concern contingent properties of the physical world that may
well either fail to hold in the quantum domain, or simply become
meaningless. If we believe that likewise Quantum Gravity
\emph{is} Quantum Geometry, we should be able to argue for it
by some sort of extension or adaptation of EEP into the
quantum domain. As a first attempt in this direction one might
ask for the status of EEP if the matter used to probe it is
described by ordinary non-relativistic Quantum Mechanics.
Can the quantum nature of matter be employed to push the bounds
on possible violations of EEP to hitherto unseen lower limits?

In this contribution I shall discuss some aspects related to this
question and, in particular, to a recent claim~\cite{Mueller.etal:2010},
according to which atom interferometric gravimeters have actually
already tested the weakest part of EEP, the universality of gravitational
redshift, and thereby improved the validity of EEP by about  four
orders of magnitude! I will come to the conclusion that this claim
is unwarranted.%
\footnote{This is the view that I expressed in my original
talk for the same reasons as those laid out here. At that time
the brief critical note \cite{Wolf.etal:2010a} and the reply
\cite{Mueller.etal:2010b} by the original proponents had
appeared in the Nature issue of September 2nd. In the meantime more
critique has been voiced \cite{Wolf.etal:2011,Samuel.Sinha:2011},
though the original claim seems to be maintained by and
large~\cite{Hohensee.etal:2011}.}  But before I do this in some detail,
I give a general discussion of the Einstein Equivalence Principle,
its separation into various sub principles and the logical
connection between them, and the import of one of these sub
principles, the Universality of Free Fall (UFF), into Quantum
Mechanics.
\section{Some background}
\label{sec:Background}

The theory of General Relativity rests on a number of hypotheses,
the most fundamental of which ensure, first of all, that gravity
can be described by a metric theory~\cite{ThorneLeeLightman:1973,Will:1993}.
Today these hypotheses are canonised in the Einstein Equivalence
Principle (EEP).\footnote{Note that EEP stands for ``the Einstein
Equivalence Principle'' and not ``Einstein's Equivalence Principle''
because Einstein never expressed it in the modern canonised form.}
EEP consists of three parts:
\subsubsection*{UFF: The Universality of Free Fall}
UFF states that free fall of ``test particles'' (further remarks on
that notion will follow) only depend on their initial position
and direction in spacetime. Hence test particles define a path
structure on spacetime in the sense of
\cite{Ehlers.Koehler:1977,Coleman.Korte:1980} which, at this
stage, need not necessarily be that of a linear connection.
In a Newtonian setting UFF states that the quotient of the
inertial and gravitational mass is a universal constant, i.e.
independent of the matter the test particle is made of.
UFF is also often called the \emph{Weak Equivalence Principle},
abbreviated by WEP, but we shall stick to the label UFF which
is more telling.

Possible violations of UFF are parametrised by the E\"otv\"os
factor, $\eta$, which measures the difference in acceleration of
two test masses made of materials $A$ and $B$:
\begin{equation}
\label{eq:DefEoetvoesFactor}
\eta(A,B)=2\cdot\frac{\vert a(A)-a(B)\vert}{a(A)+a(B)}
\approx\sum_\alpha\eta_\alpha\left(
\frac{E_\alpha(A)}{m_i(A)c^2}-\frac{E_\alpha(B)}{m_i(B)c^2}\right)\,.
\end{equation}
The second and approximate equality arises if one supposes
that violations occur in a specific fashion, for each
fundamental interaction (labelled by $\alpha$) separately.
More specifically, one expresses the gravitational mass
of the test particle made of material $A$ in terms of its
inertial mass and a sum of corrections, one for each
interaction $\alpha$, each being proportional to the
fraction that the $\alpha$'s interaction makes to the
total rest energy (cf. Sect.\,2.4 of \cite{Will:1993}):
\begin{equation}
\label{eq:EoetvoesIneractionWise}
m_g(A)=m_i(A)+\sum_\alpha \,\eta_\alpha\,\frac{E_\alpha(A)}{m_i(A)c^2}\,.
\end{equation}
Here the $\eta_\alpha$ are universal constants depending
only on the interaction but not on the test particle.
Typical numbers from modern laboratory tests, using rotating
torsion balances, are below the $10^{-12}$ level. Already in
1971 Braginsky and Panov claimed to have reached an accuracy
$\eta(\mathrm{Al},\mathrm{Pt})<9\times 10^{-13}$
for the element pair Aluminium and
Platinum~\cite{Braginsky.Panov:1972}. Currently the lowest
bound is reached for the elements Beryllium and
Titanium~\cite{Schlamminger.etal:2008}.%
\footnote{Besides for technical experimental reasons, these two
elements were chosen to maximise the difference in baryon number
per unit mass.}
 %
\begin{equation}
\label{eq:EoetvoesTests}
\eta(\mathrm{Be},\mathrm{Ti})< 2.1\times 10^{-13}\,.
\end{equation}
Resolutions in terms of $\eta_\alpha$'s of various tests are
discussed in~\cite{Will:1993}. Future tests like MICROSCOPE
(``MICRO-Satellite \`a tra\^{\i}n\'ee Compens\'ee pour l'Ob\-ser\-va\-tion du
Principe d'Equivalence'', to be launched in 2014) aim at a
lower bound of $10^{-15}$. It is expected that freely falling
Bose-Einstein condensates will also allow precision tests of UFF,
this time with genuine quantum matter~\cite{vanZoest:2010}.

\subsubsection*{LLI: Local Lorentz Invariance}
LLI states that local non-gravitational experiments exhibit no
preferred directions in spacetime, neither timelike nor spacelike.
Possible violations of LLI concern, e.g., orientation-dependent
variations in the speed of light, measured by $\Delta c/c$,
or the spatial orientation-dependence of atomic energy levels.
In experiments of the Michelson-Morley type, where $c$ is
the mean for the round-trip speed, the currently lowest
bound from laboratory experiments based on experiments with
rotating optical resonators is~\cite{Herrmann.etal:2003}:
\begin{equation}
\label{eq:MM_Herrmann}
\frac{\Delta c}{c}<3.2\times 10^{-16}\,.
\end{equation}
Possible spatial orientation-dependencies of atomic energy levels
have also been constrained by impressively low  upper bounds in
so-called Hughes-Drever type experiments.

\subsubsection*{LPI: Local Position Invariance}
LPI is usually expressed by saying that ``The outcome
of any local non-gravitational experiment is independent
of where and when in the universe it is performed''
(\cite{Will.LivingReviews:2006-3}, Sect.\,2.1). However,
in almost all discussions this is directly translated into
the more concrete \emph{Universality of Clock Rates (UCR)}
or the \emph{Universality of Gravitational redshift (UGR)},
which state that the rates of standard clocks agree if
taken along the same world line (relative comparison) and
that they show the standard redshift if taken along different
worldlines and intercompared by exchange of electromagnetic
signals. Suppose a field of light rays intersect the timelike
worldlines $\gamma_{1,2}$ of two clocks, the four-velocities of
which are $u_{1,2}$. Then the ratio of the instantaneous
frequencies measured at the intersection points of one
integral curve of the wave-vector $k$ with $\gamma_1$ and $\gamma_2$ is
\begin{equation}
\label{eq:FrequencyShiftGeneral}
\frac{\nu_2}{\nu_1}=
\frac{g(u_2,k)\vert_{\gamma_2}}{g(u_1,k)\vert_{\gamma_1}}\,.
\end{equation}
Note that this does not distinguish between gravitational and
Doppler shifts, which would be meaningless unless a local notion of
``being at rest'' were introduced. The latter requires a distinguished
timelike vector field,  as e.g. in stationary spacetimes with
Killing field $K$. Then the purely gravitational part of
(\ref{eq:FrequencyShiftGeneral}) is given in case both clocks
are at rest, i.e. $u_{1,2}=K/\Vert K\Vert\vert_{\gamma_{1,2}}$,
where $\gamma_{1,2}$ are now two different integral lines of $K$ and
$\Vert K\Vert:=\sqrt{g(K,K)}$:
\begin{equation}
\label{eq:FrequencyShiftStationary}
\frac{\nu_2}{\nu_1}:=\frac%
{g\bigl(k,K/\Vert K\Vert\bigr)\vert_{\gamma_2}}%
{g\bigl(k,K/\Vert K\Vert\bigr)\vert_{\gamma_1}}
=\sqrt{\frac{g(K,K)\vert_{\gamma_1}}{g(K,K)\vert_{\gamma_2}}}\,.
\end{equation}
The last equality holds since $g(k,K)$ is constant along
the integral curves of $k$, so that $g(k,K)\vert_{\gamma_1}
=g(k,K)\vert_{\gamma_2}$ in (\ref{eq:FrequencyShiftStationary}),
as they lie on the same integral curve of $k$. Writing
$g(K,K)=:1+2U/c^2$ and assuming $U/c^2\ll 1$, we get
\begin{equation}
\label{eq:FrequencyShiftStationary-Koord}
\frac{\Delta\nu}{\nu}:=\frac{\nu_2-\nu_1}{\nu_1}
=-\frac{U_2-U_1}{c^2}\,.
\end{equation}
Possible deviations from this result are usually parametrised
by multiplying the right-hand side of
(\ref{eq:FrequencyShiftStationary-Koord}) with $(1+\alpha)$,
where $\alpha=0$ in GR. In case of violations of UCR/UGR, $\alpha$
may depend on the space-time point and/or on the type of
clock one is using. The lowest upper bound on $\alpha$ to date
for comparing (by electromagnetic signal exchange) clocks on
\emph{different} worldlines derives from an experiment made in
1976 (so-called ``Gravity Probe A'') by comparing a hydrogen-maser
clock in a rocket, that during a total experimental time of 1~hour
and 55~minutes was boosted to an altitude of about $10\,000\,\mathrm{km}$,
to a similar clock on the ground. It led to~\cite{Vessot.etal:1980}
\begin{equation}
\label{eq:UpperBoundAlphaRS}
\alpha_{\rm RS}<7\times 10^{-5}\,.
\end{equation}
The best relative test, comparing different clocks
(a ${}^{199}\mathrm{Hg}$-based optical clock and one based on
the standard hyperfine splitting of ${}^{133}\mathrm{Cs}$)
along the (almost) \emph{same} worldline for six years
gives~\cite{Fortier.etal:2007}
\begin{equation}
\label{eq:UpperBoundAlphaCR}
\alpha_{\rm CR}<5.8\times 10^{-6}\,.
\end{equation}
Here and above ``RS'' and ``CR'' refer to ``redshift'' and
``clock rates'', respectively, a distinction that we prefer to
keep from now on in this paper, although it is not usually
made. To say it once more: $\alpha_{\rm RS}$ parametrises possible
violations of UGR by comparing identically constructed clocks
moving along different worldlines, whereas $\alpha_{\rm CR}$
parametrises possible violations of UCR by comparing clocks of
different construction and/or composition moving more or less
on the same worldline. An improvement in putting upper bounds on
$\alpha_{\rm CR}$, aiming for at least $2\times 10^{-6}$, is expected
from ESA's ACES mission (ACES = Atomic Clock Ensemble in Space),
in which a Caesium clock and a H-maser clock will be flown to
the Columbus laboratory at the International Space Station (ISS),
where they will be compared for about two
years~\cite{Cacciapuoti.Salomon:2009}.

\begin{rem}
\label{rem:TestParticles}
The notion of ``test particles'' essentially used in the formulation
of UFF is not without conceptual dangers. Its intended meaning is
that of an object free of the ``obvious'' violations of UFF,
like higher multipole moments in its mass distribution
and intrinsic spin (both of which would couple to the spacetime
curvature) and electric charge (in order to avoid problems
with radiation reaction). Moreover, the test mass should
not significantly back-react onto the curvature of
spacetime and should not have a significant mass defect
due to its own gravitational binding. It is clear that the
simultaneous fulfilment of these requirements will generally
be context-dependent. For example, the Earth will count
with reasonable accuracy as a test particle as far as its
motion in the Sun's gravitational field is concerned, but
certainly not for the Earth-Moon system. Likewise, the
notion of ``clock'' used in UCR/UGR intends to designate
a system free of the ``obvious'' violations. In GR a
``standard clock'' is any system that allows to measure the
length of timelike curves. If the curve is accelerated it is
clear that some systems cease to be good clocks (pendulum clocks)
whereas others are far more robust. An impressive example for
the latter is muon decay, where the decay time is affected by a fraction
less than $10^{-25}$ at an acceleration of $10^{18}g$~\cite{Eisele:1987}.
On the other hand, if coupled to an accelerometer, eventual
disturbances could  in principle always be corrected for. At
least as far as classical physics is concerned, there seems to be
no serious lack of real systems that classify as test particles and
clocks in contexts of interest. But that is a contingent
property of nature that is far from self-evident.
\end{rem}

\begin{rem}
The lower bounds for UFF, LLI, and UCR/UGR quoted above
impressively show how much better UFF and
LLI are tested in comparison to  UCR/UGR. This makes the latter
the weakest member in the chain that constitutes EEP. It would
therefore be desirable to significantly lower the upper bounds
for violations of the latter. Precisely this has recently
(February 2010) been claimed in~\cite{Mueller.etal:2010} by
remarkable four orders in magnitude - and without doing a single
new experiment! This will be analysed in detail below.
\end{rem}

It can be carefully argued for (though not on the level
of a mathematical theorem) that only metric theories can
comply with EEP. In particular, the additional requirements
in EEP imply that the path structure implied by UFF alone
must be that of a linear connection. Metric theories, on
the other hand, are defined by the following properties
(we state them with slightly different wordings as compared
to \cite{Will.LivingReviews:2006-3}, Sect.\,2.1):
\begin{itemize}
\item[M1.]
Spacetime is a four-dimensional differentiable manifold,
which carries a metric (symmetric non-degenerate bilinear form)
of Lorentzian signature, i.e. $(-,+,+,+)$ or $(+,-,-,-)$,
depending on convention%
\footnote{Our signature convention will be the
``mostly minus'' one, i.e. $(+,-,-,-)$.}.
\item[M2.]
The trajectories of freely falling test bodies are geodesics
of that metric.
\item[M3.]
With reference to freely falling frames, the non-gravitational
laws of physics are those known from Special Relativity.
\end{itemize}

This canonisation of EEP is deceptive insofar as it suggests
an essential logical independence of the individual hypotheses.
But that is far from true. In fact, in 1960 the surprising
suggestion has been made by Leonard Schiff that UFF should
imply EEP, and that hence UFF and EEP should, in fact, be
equivalent; or, expressed differently, UFF should already
imply LLI and LPI. This he suggested in a ``note added in proof''
at the end of his classic paper \cite{Schiff:1960}, in the
body of which he asked the important question whether the
three classical ``crucial tests'' of GR were actually sensitive
to the precise form of the field equations (Einstein's equations)
or whether they merely tested the more general equivalence
principle. He showed that the gravitational red-shift and the
deflection of light could be deduced from EEP and that only the
correct evaluation of the precession of planetary orbits needed
an input from Einstein's equations. If true, it follows that any
discrepancy between theory and experiment would have to be
reconciled with the experimentally well-established validity
of the equivalence principle and Special Relativity. Hence Schiff
concludes:
\begin{quote}
``By the same token, it will be extremely difficult to design
a terrestrial or satellite experiment that really tests general
relativity, and does not merely supply corroborative evidence
for the  equivalence principle [meaning UFF; D.G.] and special
relativity. To accomplish this it will be necessary either to
use particles of finite rest mass so that the geodesic equation
may be confirmed beyond the Newtonian approximation, or to verify
the exceedingly small time or distance changes of order
$(GM/c^2r)^2$. For the latter the required accuracy of a clock
is somewhat better than one part in $10^{18}$.''
\end{quote}
Note that this essentially says that testing GR means foremost
to test UFF, i.e. to perform E\"otv\"os-type experiments.

This immediately provoked a contradiction by Robert Dicke in
\cite{Dicke:1960}, who read Schiff's assertions as ``serious
indictment of the very expensive government-sponsored program
to put an atomic clock into an artificial satellite''. For, he
reasoned, ``If Schiff's basic assumptions are as firmly established
as he believes, then indeed this project is a waste of government
funds.'' Dicke goes on to point out that for several reasons
UFF is not as well tested by past E\"otv\"os-type experiments
as Schiff seems to assume and hence argues in strong favour of
the said planned tests.

As a reaction to Dicke, Schiff added in proof the justifying
note already mentioned above. In it he said:
\begin{quote}
``The E\"otv\"os experiments show with considerable accuracy that
the gravitational and inertial masses of normal matter are equal.
This means that the ground-state eigenvalue of the Hamiltonian
for this matter appears equally in the inertial mass and in the
interaction of this mass with a gravitational field. It would be
quite remarkable if this could occur without the entire
Hamiltonian being involved in the same way, in which case a clock
composed of atoms whose motions are determined by this Hamiltonian
would have its rate affected in the expected manner by a
gravitational field.''
\end{quote}
This is the origin of what is called \emph{Schiff's conjecture} in the
literature. Attempts have been made to ``prove'' it in special
situations \cite{LightmanLee:1973}, but it is well known not to
hold in mathematical generality. For example, consider gravity and
electromagnetism coupled to point charges just as in GR, but now
make the single change that the usual Lagrangian density
$-\frac{1}{4}F_{ab}F^{ab}$ for the free electromagnetic
field is replaced by $-\frac{1}{4}C^{ab\,cd}F_{ab}F_{cd}$, where
the tensor field $C$ (usually called the constitutive tensor; it
has the obvious symmetries of the Riemann tensor) can be any
function of the metric. It is clear that this change implies
that for general $C$ the laws of (vacuum) electrodynamics in a
freely falling frame will not reduce to those of Special
Relativity and that, accordingly, Schiff's conjecture cannot
hold for all $C$. In fact, Ni proved \cite{Ni:1977} that Schiff's
conjecture holds iff
\begin{equation}
\label{eq:NiConstitutiveTensor}
C^{ab\,cd}=\tfrac{1}{2}\bigl(g^{ac}g^{bd}-g^{ad}g^{bc}\bigr)
+\phi\varepsilon^{abcd}\,,
\end{equation}
where $\phi$ is some scalar function of the metric.

Another and simpler reasoning, showing that UFF cannot by itself
imply that gravity is a metric theory in the semi-Riemannian sense
(rather than, say, of Finslerian type) is the following: Imagine
the ratio of electric charge and inertial mass were a universal
constant for all existing matter and that a fixed electromagnetic
field existed throughout spacetime. Test particles would move
according to the equation
\begin{equation}
\label{eq:ForcedGeodesic}
{\ddot x}^a+\Gamma^a_{bc}{\dot x}^b{\dot x}^c=(q/m)F^a_b{\dot x}^b\,,
\end{equation}
where the $\Gamma$'s are the Christoffel symbols for the metric
and $(q/m)$ is the said universal constant. This set of four
ordinary differential equations for the four functions $x^a$ clearly
defines a path structure on spacetime, but for a general $F_{ab}$ there
will be no semi-Riemannian metric with respect to which
(\ref{eq:ForcedGeodesic}) is the equation for a geodesic.
Hence Schiff's conjecture should at best be considered
as a selection criterion.

\subsection{LLI and UGR}
We consider a static homogeneous and downward-pointing
gravitational field $\vec g=-g\vec e_z$. We follow
Section\,2.4 of \cite{Will:1993} and assume the
validity of UFF and LLI but allow for violations of
LPI. Then UFF guarantees the local existence of a
freely-falling frame with coordinates $\{x_f^\mu\}$,
whose acceleration is the same as that of test particles.
For a rigid acceleration we have
\begin{alignat*}{2}
&ct_f&&\,=\,(z_s+c^2/g)\,\sinh(gt_s/c)\,,\\
&x_f&&\,=\,x_s\,,\\
&y_f&&\,=\,y_s\,,\\
&z_f&&\,=\,(z_s+c^2/g)\,\coth(gt_s/c)\,.
\end{alignat*}
LLI guarantees that, \emph{locally}, time measured by, e.g.,
an atomic clock is proportional to Minkowskian proper length in
the freely falling frame. If we consider violations of LPI,
the constant of proportionality might depend on the space-time
point, e.g. via dependence on the gravitational potential $\phi$,
as well as the type of clock:
\begin{alignat}{2}
c^2\,d\tau^2
&\,=\,F^2(\phi)\bigl[c^2dt_f^2-dx_f^2-dy_f^2-dz_f^2\bigr]\\
&\,=\,F^2(\phi)\left[\left(1+\frac{gz_s}{c^2}\right)^2c^2dt_s^2
-dx_s^2-dy_s^2-dz_s^2\right]\,.
\end{alignat}
The \emph{same} time interval $dt_s=dt_s(z_s^{(1)})=dt_s(z_s^{(2)})$
on the two static clocks at rest wrt. $\{x^\mu_s\}$, placed at different
heights $z_s^{(1)}$ and $z_s^{(2)}$, correspond to \emph{different}
intervals $d\tau^{(1)},d\tau^{(2)}$ of the inertial clock, giving
rise to the redshift (all coordinates are $\{x^\mu_s\}$ now,
so we drop the subscript $s$):
\begin{equation}
\zeta:=\frac{d\tau^{(2)}-d\tau^{(1)}}{d\tau^{(1)}}
=\frac{F(z^{(2)})(1+gz^{(2)}/c^2)}{F(z^{(1)})(1+gz^{(1)}/c^2)}\ -1 \,.
\end{equation}
For small $\Delta z=z^{(2)}-z^{(1)}$ this gives to first order in
$\Delta z$
\begin{equation}
\Delta\zeta=(1+\alpha)g\Delta z/c^2 \,,
\end{equation}
where
\begin{equation}
\alpha=\frac{c^2}{g}\ \bigl(\vec e_z\cdot\vec\nabla\ln(F)\bigr)
\end{equation}
parametrises the deviation from the GR result. $\alpha$ may depend on
position, gravitational potential, and the type of clock one is
using.

\subsection{Energy conservation, UFF, and UGR}
\label{sec:EnergyCons-UFF_UGR}
\begin{figure}[bht]
\centering
\includegraphics[width=0.583\linewidth]{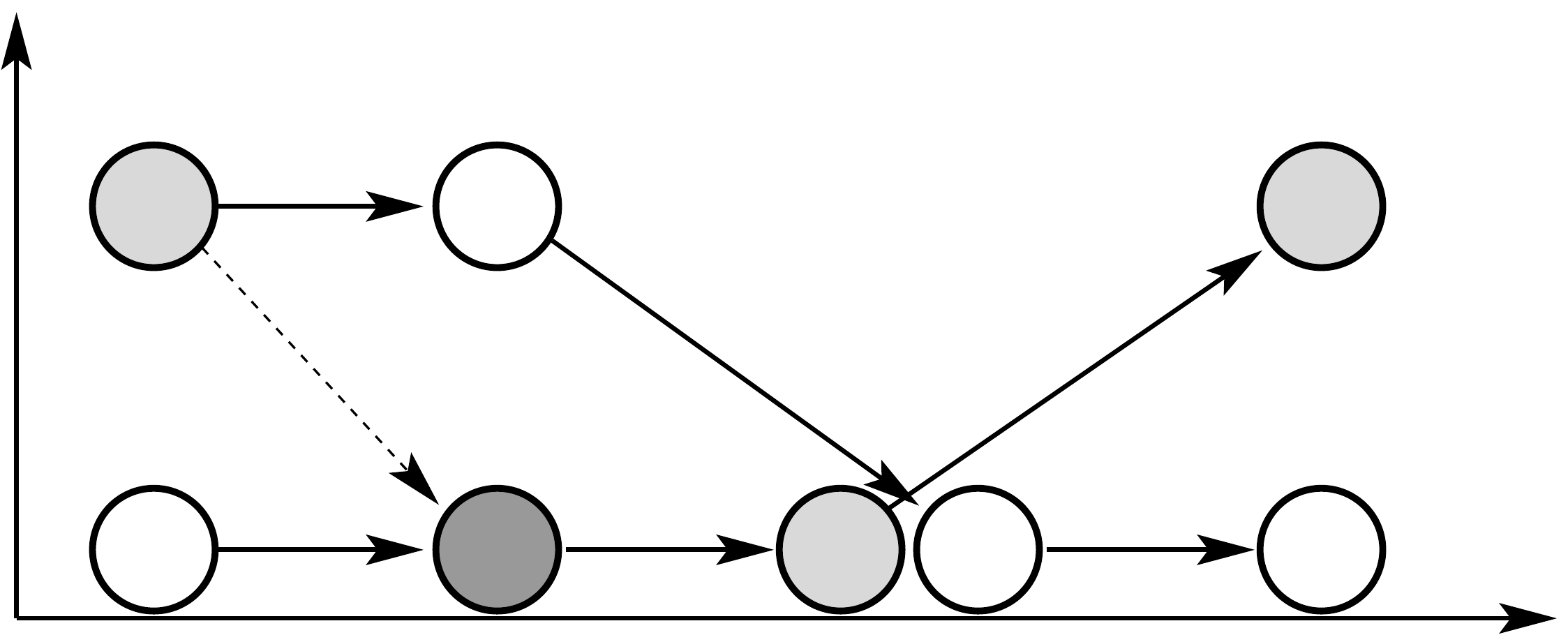}
\put(3,-1){\footnotesize time}
\put(-180,-8){$T_1$}
\put(-137,-8){$T_2$}
\put(-85,-8){$T_3$}
\put(-33,-8){$T_4$}
\put(-153,32){\footnotesize $h\nu$}
\put(-103,35){\footnotesize $g_A$}
\put(-68,35){\footnotesize $g_B$}
\put(-219,72){\footnotesize height}
\put(-179,8){\footnotesize $A$}
\put(-178.5,50){\footnotesize $B$}
\put(-136.6,8){\footnotesize $B'$}
\put(-136,50){\footnotesize $A$}
\put(-94,8){\footnotesize $B$}
\put(-76.4,8){\footnotesize $A$}
\put(-34,8){\footnotesize $A$}
\put(-34,50){\footnotesize $B$}
\caption{\label{fig:NordtvedtGedExp} Nordtvedt's gedanken experiment.
Two systems in three different energy states $A$, $B$, and $B'$ are
considered at four different times $T_1, T_2, T_3$, and $T_4$.
The initial state at $T_1$ and the final state at $T_4$ are identical,
which, by means of energy conservation, leads to an interesting
quantitative relation between possible violations of UFF and UGR.}
\end{figure}
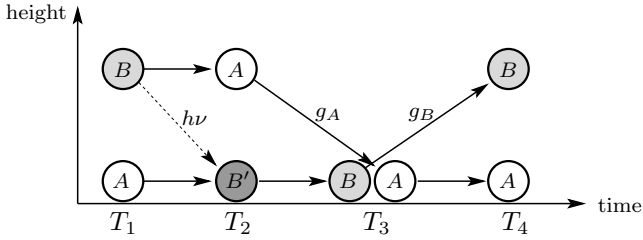

In this subsection we wish to present some well-known
gedanken-experiment-type arguments \cite{Nordtvedt:1975,Haugan:1979}
according to which there is a link between violations of
UFF and UGR, provided energy conservation holds. Here we
essentially present Nordtvedt's version; compare Figure\,\ref{fig:NordtvedtGedExp}.

We consider two copies of a system that is capable of 3 energy
states $A,B$, and $B'$ (white, light grey, grey), with $E_A<E_B<E_{B'}$,
placed into a vertical downward-pointing homogeneous gravitational
field. Initially system~2 is in state $B$ and placed at height $h$
above system~1, which is in state $A$. At time $T_1$ system 2 makes
a transition $B\rightarrow A$ and sends out a photon of energy
$h\nu=E_B-E_A$. At time $T_2$ system 1 absorbs this photon, which
is now blue-shifted due to its free fall in the downward-pointing
gravitational field, and makes a transition $A\rightarrow B'$.
At $T_3$ system 2 has been dropped from height $h$ with an
acceleration of modulus $g_A$ that possibly depends on its
inner state $A$ and has hit system 1 inelastically, leaving
one system in state $A$ and at rest, and the other system in
state $B$ with an upward motion. By energy conservation this
upward motion has a kinetic energy of
\begin{equation}
 \label{eq:NordtvedtExp1}
E_{\rm kin}=M_Ag_Ah+(E_{B'}-E_B)\,.
\end{equation}
This upward motion is a free fall in a gravitational field
and since the system is now in an inner state $B$, it is
decelerated with modulus $g_B$, which may differ from $g_A$.
At $T_4$ the system in state $B$ has climbed to the
height $h$, which must be the same as the height at the
beginning, again by energy conservation. Hence we have
$E_{\rm kin}=M_Bg_Bh$, and since moreover
$E_{B'}-E_B=(M_{B'}-M_B)c^2$, we get
\begin{equation}
\label{eq:NordtvedtExp2}
M_Ag_Ah+M_{B'}c^2=M_Bc^2+M_Bg_Bh\,.
\end{equation}
Therefore
\begin{equation}
\label{eq:NordtvedtRelBetaEP-a}
\begin{split}
\frac{\delta\nu}{\nu}
&:=\frac{(M_{B'}-M_A)-(M_B-M_A)}{M_B-M_A}\\
&=\frac{g_Bh}{c^2}\left[1+\frac{M_A}{M_B-M_A}
\frac{g_B-g_A}{g_B}\right]\,,
\end{split}
\end{equation}
so that
\begin{equation}
\label{eq:NordtvedtRelBetaEP-b}
\alpha=\frac{M_A}{M_B-M_A}\frac{g_B-g_A}{g_B}
=:\frac{\delta g/g}{\delta M/M}\,.
\end{equation}

This equation gives a quantitative link between violations of
UFF, here represented by $\delta g$, and violations of UGR,
here represented by $\alpha$. The strength of the link depends
on the fractional difference of energies/masses $\delta M/M$,
which varies according to the type of interaction that is
responsible for the transitions. Hence this equation can answer
the question of how accurate a test of UGR must be in order to
test the metric nature of gravity to the same level of accuracy
than E\"otv\"os-type experiments. Given that for the latter we have
$\delta g/g<10^{-13}$, this depends on the specific situation
(interaction) through $\delta M/M$. For atomic clocks the
relevant interaction is the magnetic one, since the energies
rearranged in hyperfine transitions correspond to magnetic
interactions. The variation of the magnetic contribution to
the overall self-energy between pairs of chemical elements
$(A,B)$ for which the E\"otv\"os factor has been strongly bounded
above, like Aluminium and Gold (or Platinum), have been (roughly)
estimated to be $\vert\delta M/M\vert\approx 5\times 10^{-5}$
\cite{Nordtvedt:1975}. Hence one needed a precision of
$\alpha_{\rm RS}<5\times 10^{-9}$ for a UGR test to match existing
UFF tests.

\section[UFF in quantum mechanics]{UFF in Quantum Mechanics}
\label{sec:UffInQM}
In classical mechanics the universality of free fall is
usually expressed as follows: We consider Newton's Second
Law for a point particle of inertial mass $m_i$,
\begin{equation}
\label{eq:NewtonsEq}
\vec F=m_i\ddot{\vec x}\,,
\end{equation}
and specialise it to the case in which the external force is
gravitational, i.e. $\vec F=\vec F_{\rm grav}$, where
\begin{equation}
\label{eq:GravForce}
\vec F=m_g\,\vec g\,.
\end{equation}
Here $m_g$ denotes the passive gravitational mass and
$\vec g:\mathbb{R}^4\rightarrow\mathbb{R}^3$ is the
(generally space and time dependent) gravitational field.
Inserting (\ref{eq:GravForce}) into (\ref{eq:NewtonsEq}) we get
\begin{equation}
\label{eq:NewtonsEqOfMotion}
\ddot{\vec x}(t)=
\left(\frac{m_g}{m_i}\right)\,\vec g\bigl(t,\vec x(t)\bigr)\,.
\end{equation}
Hence  the solution of (\ref{eq:NewtonsEqOfMotion}) only depends
on the initial time,  spatial position,  and spatial velocity
iff $m_g/m_i$ is a universal constant, which by appropriate choices
of units can be made unity.

This reasoning is valid for point particles only. But it clearly
generalises to the centre-of-mass motion of an extended mass
distribution in case of spatially homogeneous gravitational fields,
where $m_i$ and $m_g$ are then the total inertial and total
(passive) gravitational masses. For this generalisation to hold
it need not be the case that the spatial distributions of inertial
and gravitational masses are proportional. If they are not
proportional, the body will deform as it moves under the
influence of the gravitational field. If they are proportional
and the initial velocities of all parts of the body are the
same, the trajectories of the parts will all be translates of
one another. If the initial velocities are not the same, the
body will disperse without the action of internal cohesive forces
in the same way as it would without gravitational field.

There is no pointlike-supported wave packet in quantum
mechanics. Hence we ask for the analogy to the situation just
described: How does a wave packet fall in a homogeneous
gravitational field? The answer is given by the following result,
the straightforward proof of which we suppress.
\begin{prop}
$\psi$ solves the Schr\"odinger Equation
\begin{equation}
\label{eq:SchroedingerEqHomogForce}
i\hbar\partial_t\psi=
\left(-\frac{\hbar^2}{2m_i}\Delta-\vec F(t)\cdot\vec x\right)\psi
\end{equation}
iff
\begin{equation}
\label{eq:TransPsi}
\psi=\bigl(\exp(i\alpha)\,\psi'\bigr)\circ\Phi^{-1} \,,
\end{equation}
where $\psi'$ solves the free Schr\"odinger equation (i.e. without
potential). Here $\Phi:\mathbb{R}^4\rightarrow\mathbb{R}^4$ is the
following spacetime diffeomorphism (preserving time)
\begin{equation}
\label{eq:SpaceTimeDiffeo}
\Phi(t,\vec x)=\bigl(t,\vec x+\vec\xi(t)\bigr)\,,
\end{equation}
where $\vec\xi$ is a solution to
\begin{equation}
\label{eq:SpaceTimeDiffeo-spatial}
\ddot{\vec\xi}(t)=\vec F(t)/m_i
\end{equation}
with $\vec\xi(0)=\vec 0$, and $\alpha:\mathbb{R}^4\rightarrow\mathbb{R}$
given by
\begin{equation}
\label{eq:AlphaFunction}
\alpha(t,\vec x)=\frac{m_i}{\hbar}
\left\{
\dot{\vec\xi}(t)\cdot\bigl(\vec x+\vec\xi(t)\bigr)-
\frac{1}{2}\int^tdt'\Vert\dot{\vec\xi}(t')\Vert^2
\right\}\,.
\end{equation}
\end{prop}

To clearly state the simple meaning of (\ref{eq:TransPsi})
we first remark that changing a trajectory
$t\mapsto\vert\psi(t)\rangle$ of Hilbert-space vectors
to $t\mapsto\exp\bigl(i\alpha(t)\bigr)\,\vert\psi(t)\rangle$
results in the \emph{same} trajectory of states, since
the state at time $t$ is faithfully represented by the
ray in Hilbert space generated by the vector (unobservability
of the global phase). As our Hilbert space is that of square-integrable functions on $\mathbb{R}^2$, only the $\vec x$-dependent parts of the phase (\ref{eq:AlphaFunction})
change the instantaneous state. Hence, in view of
\eqref{eq:AlphaFunction}, the meaning of \eqref{eq:TransPsi}
is that the state $\psi$ at time $t$ is obtained from
the freely evolving state at time $t$ with the same initial
data by 1)~a boost with velocity $\vec v=\dot{\vec\xi}(t)$
and 2)~a spatial displacement by $\vec\xi(t)$.
In particular, the spatial probability distribution
$\rho(t,\vec x):=\psi^*(t,\vec x)\psi(t,\vec x)$ is of
the form
\begin{equation}
\label{eq:ProbDist}
\rho=\rho'\circ\Phi^{-1} \,,
\end{equation}
where $\rho'$ is the freely evolving spatial probability
distribution. This implies that the spreading of $\rho$
is entirely that due to the free evolution.

Now specialise to a homogeneous and static gravitational
field $\vec g$, such that $\vec F=m_g\vec g$; then
\begin{equation}
\label{eq:XiForHomStatField-a}
\vec\xi(t)=\vec vt+\tfrac{1}{2}\vec a t^2
\end{equation}
with
\begin{equation}
\label{eq:XiForHomStatField-b}
\vec a=(m_g/m_i)\,\vec g\,.
\end{equation}
In this case the phase (\ref{eq:AlphaFunction}) is

\begin{equation}
\label{eq:AlphaForHomStatField-a}
\alpha(t,\vec x)=\frac{m_i}{\hbar}
\Bigl\{
\vec v\cdot\vec x+
\bigl(\tfrac{1}{2}v^2+\vec a\cdot\vec x\bigr)\,t
+\vec v\cdot\vec a\,t^2+\tfrac{1}{3}a^2t^3
\Bigr\}\,,
\end{equation}
where $v:=\Vert\vec v\Vert$ and $a:=\Vert\vec a\Vert$.

As the spatial displacement $\vec\xi(t)$ just depends on
$m_g/m_i$, so does that part of the spatial evolution
of $\rho$ that is due to the interaction with the
gravitational field. This is a quantum-mechanical
version of UFF. Clearly, the inevitable spreading of
the free wave packet, which depends on $m_i$ alone,
is just passed on to the solution in the gravitational
field. Recall also that the evolution of the full state
involves the $\vec x$-dependent parts of the phase,
which correspond to the gain in momentum during free
fall. That gain due to acceleration is just the
classical $\delta\vec p=m_i\vec at$ which, in view of
(\ref{eq:XiForHomStatField-b}), depends on $m_g$ alone.

Other dependencies of physical features on the pair
$(m_g,m_i)$ are also easily envisaged. To see this, we
consider the stationary case of (\ref{eq:SchroedingerEqHomogForce}),
where $i\hbar\partial_t$ is replaced by the energy $E$, and
also take the external force to correspond to a constant
gravitational field in negative $z$-direction: $\vec F=-m_gg\vec e_z$.
The Schr\"odinger equation then separates, implying free motion
perpendicular to the $z$-direction. Along the $z$-direction
one gets
\begin{equation}
\label{eq:Airy-a}
\left(\frac{d^2}{d\zeta^2}-\zeta\right)\,\psi=0,
\end{equation}
with
\begin{equation}
\label{eq:Airy-b}
\zeta:=\kappa z-\varepsilon\,,
\end{equation}
where
\begin{equation}
\label{eq:Airy-c}
\kappa:=\left[\frac{2m_im_gg}{\hbar^2}\right]^{\frac{1}{3}}\,,\quad
\varepsilon:=E\cdot\left[\frac{2m_i}{m_g^2g^2\hbar^2}\right]^{\frac{1}{3}}\,.
\end{equation}

\begin{figure}[h]
\centering
\includegraphics[width=0.583\linewidth]{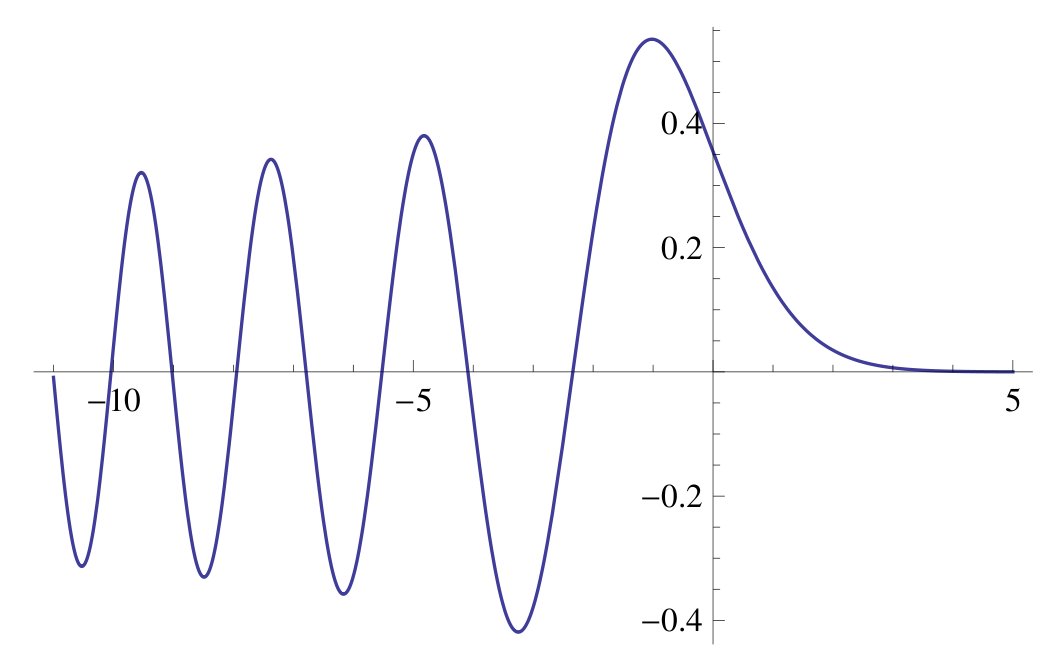}
\put(2,52){\footnotesize $\zeta$}
\put(-70,125){\footnotesize $\mathrm{Ai}(\zeta)$}
\caption{\label{fig:AiryPlot}
Airy function from $\zeta=-10$ to $\zeta=5$.}
\end{figure}

A solution to (\ref{eq:Airy-a}) that falls off for
$\zeta\rightarrow\infty$ must be  proportional to the
Airy function, a plot of which is shown in
Figure\,\ref{fig:AiryPlot}.

As has been recently pointed out in \cite{Kajari.etal:2010},
it is remarkable that the penetration depth into the classically
forbidden region, which is a simple function of the length
$\kappa^{-1}$, depends on the \emph{product} of inertial
and gravitational mass. Also, suppose we put an infinite potential
barrier at $z=0$. Then the energy eigenstates of a particle in the
region $z>0$ are obtained by the requirement $\psi(0)=0$, hence
$\varepsilon=-z_n$, where  $z_n<0$ is the $n$-th zero of the
Airy function. By (\ref{eq:Airy-c}) this gives
\begin{equation}
\label{eq:Airy-d}
E_n=\left[\frac{m^2_gg^2\hbar^2}{2m_i}\right]^\frac{1}{3}\cdot (-z_n)\,.
\end{equation}
The energy eigenvalues of this ``atom trampoline'' \cite{Kajari.etal:2010}
depend on the combination $m_g^2/m_i$. In the gravitational field of the
Earth the lowest-lying  energies have been realised with ultracold
neutrons \cite{Nesvizhevsky:2002}; these energies are just a few
$10^{-12}\,\mathrm{eV}$.

In classical physics, the return-time of a body that is projected
at level $z=0$ against the gravitational field $\vec g=-g\vec e_z$
in positive $z$-direction, so that it reaches a maximal height of
$z=h$, is given by
\begin{equation}
\label{eq:ReturnTime_Classical}
T_{\rm ret}=2\cdot\left[\frac{m_i}{m_g}\right]^\frac{1}{2}\cdot
\left[\frac{2h}{g}\right]^\frac{1}{2}\,.
\end{equation}
Now, in Quantum Mechanics we may well expect the return time to
receive corrections from barrier-penetration effects, which one
expects to delay arrival times. Moreover, since the penetration
depth is a function of the product rather than the quotient of
$m_g$ and $m_i$, this correction can also be expected to introduce
a more complicated dependence of the return time on the two
masses. It is therefore somewhat surprising to learn that
the classical formula (\ref{eq:ReturnTime_Classical}) can be
reproduced as an exact quantum-mechanical result~\cite{Davies:2004}.
This is not the case for other shapes of the potential. A simple
calculation confirms the intuition just put forward for a step
potential, which leads to a positive correction to the classical
return which is proportional to the quantum-mechanical penetration
depth and also depends on the inertial mass. Similar things can be
said of an exponential potential (see \cite{Davies:2004} for details).
Clearly, these results make no more proper physical sense than
the notion of \emph{timing} that is employed in these calculations.
This is indeed a subtle issue which we will not enter. Suffice it
to say that \cite{Davies:2004} uses the notion of a ``Peres clock''
\cite{Peres:1980} which is designed to register and store times
of flight without assuming localised particle states.
The intuitive reason why barrier penetration does not lead to
delays in return time for the linear potential may be read off
Figure\,\ref{fig:AiryPlot}, which clearly shows that the Airy
function starts decreasing \emph{before} the classical turning
point ($\zeta=0$) is reached. This has been interpreted as
saying that there is also a finite probability that the particle
is back-scattered \emph{before} it reaches the classical turning
point. Apparently this just cancels the opposite effect from
barrier penetration in case of the linear potential, thus giving
rise to an unexpectedly close analogue of UFF in Quantum Mechanics.

\section[Phase-shift calculation in non-relativistic quantum mechanics]{Phase-shift calculation in non-relativistic Quantum Mechanics}
\label{sec:PhaseShiftCalc}
We consider the motion of an atom in a static homogeneous
gravitational field $\vec g=-g\vec e_z$. We restrict attention
to the motion of the centre of mass along the $z$-axis, the
velocity of which we assume to be so slow that the Newtonian
approximation suffices. The centre of mass then obeys a simple
Sch\"odinger equation in a potential that depends linearly on
the centre-of-mass coordinate. This suggests to obtain (exact)
solutions of the time-dependent Schr\"odinger equation by using
the path-integral method; we will largely follow~\cite{Storey.CohenTannoudji:1994}.

The time evolution of a Schr\"odinger wave function in position
representation is given by
\begin{equation}
\label{eq:TimeEvolution}
\psi(z_b,t_b)=\int_{\rm space} dz_a\,
K(z_b,t_b\,;\,z_a,t_a)\,\psi(z_a,t_a)\,,
\end{equation}
where
\begin{equation}
\label{eq:Propagator}
K(z_b,t_b\,;\,z_a,t_a):=
\langle z_b\vert\exp\bigl(-iH(t_b-t_a)/\hbar\bigr)\vert z_a\rangle\,.
\end{equation}
Here $z_a$ and $z_b$ represent the initial and final position
in the vertical direction. The path-integral representation of the
propagator is
 \begin{equation}
\label{eq:PropagatorPathIntegral}
K(z_b,t_b\,;\,z_a,t_a)=
\int_{\Gamma(a,b)}\mathcal{D}z(t)\,\exp\bigl(iS[z(t)]/\hbar\bigr)\,,
\end{equation}
where
\begin{equation}
\Gamma(a,b):=
\bigl\{
z:[t_a,t_b]\rightarrow M\mid z(t_{a,b})=z_{a,b}
\bigr\}
\end{equation}
contains all continuous paths. The point is that the  path integral's
dependence on the initial and final positions $z_a$ and $z_b$ is easy
to evaluate whenever the Lagrangian is of at most
quadratic order in the positions and their velocities:
\begin{equation}
\label{eq:QuadraticLagrangian}
L(z,\dot z)=a(t)\dot z^2+b(t)\dot zz+c(t)z^2+d(t)\dot z+e(t)z+f(t)\,.
\end{equation}
Examples are: 1)~The free particle, 2)~particle in a homogeneous
gravitational field, 3)~particle in a rotating frame of reference.

To see why this is true, let $z_*\in\Gamma(a,b)$ denote the
solution to the classical equations of motion:
\begin{equation}
\label{eq:ClassicalSolutionCondition}
\frac{\delta S}{\delta z(t)}\bigg\vert_{z(t)=z_*(t)}=0\,.
\end{equation}
We parametrise an arbitrary path $z(t)$ by its difference to the
classical solution path; that is, we write
\begin{equation}
\label{eq:RelativePath}
z(t)=z_*(t)+\xi(t)
\end{equation}
and regard $\xi(t)$ as path variable:
\begin{equation}
\label{eq:PropagatorInNewVariable}
K(z_b,t_b\,;\,z_a,t_a)=
\int_{\Gamma(0,0)}\mathcal{D}\xi(t)\,
  \exp\bigl(iS[z_*+\xi]/\hbar\bigr)\,.
\end{equation}
Taylor expansion around $z_*(t)$ for each value of $t$,
taking into account~(\ref{eq:ClassicalSolutionCondition}),
gives
\begin{equation}
\begin{split}
K(z_b,t_b\,;\,z_a,t_a)=
&\exp\left\{\frac{i}{\hbar}S_*(z_b,t_b\,;\,z_a,t_a)\right\}\\
\times \int_{\Gamma(0,0)}\mathcal{D}\xi(t)
&\exp\left\{\frac{i}{\hbar}\int_{\Gamma(0,0)}dt\,\bigl[a(t){\dot\xi}^2
+b(t)\dot\xi\xi+c(t)\xi^2\bigr]\right\}\,.\\
\end{split}
\end{equation}
Therefore, for polynomial Lagrangians of at most
quadratic order, the propagator has the exact representation
\begin{equation}
\label{eq:ExactPropagator}
K(z_b,t_b\,;\,z_a,t_a)=
F(t_b,t_a)\,
\exp\left\{\frac{i}{\hbar}\,S_*(z_b,t_b\,;\,z_a,t_a)\right\} \,,
\end{equation}
where $F(t_b,t_a)$ does not depend on the initial and
final position and $S_*$ is the action for the extremising
path (classical solution). We stress once more that
(\ref{eq:ExactPropagator}) is valid only for Lagrangians
of at most quadratic order. Hence we may use it to
calculate the exact phase change for the non-relativistic
Schr\"odinger equation in a static and homogeneous
gravitational field.

\section{Free fall in a static homogeneous gravitational field}
We consider an atom in a static and homogeneous gravitational field
$\vec g=-g\vec e_z$. We restrict attention to its centre-of-mass
wave function, which we represent as that of a point particle.
During the passage from the initial to the final location the atom
is capable of assuming different internal states. These changes
will be induced by laser interaction and will bring about changes in
the inertial and gravitational masses, $m_i$ and $m_g$. It turns
out that for the situation considered here (hyperfine-split ground
states of Caesium) these changes will be negligible, as will be
shown in footnote\,\ref{foot:StateChange}. Hence we can model
the situation by a point particle of fixed inertial and gravitational
mass, which we treat as independent parameters throughout. We
will nowhere assume $m_i=m_g$.

\subsection{Some background from GR}
\label{sec:GR-Story}
In General Relativity the action for the centre-of-mass motion
for the atom (here treated as point particle) is $(-m^2c)$
times the length functional, where $m$ is the mass
(here $m_i=m_g=m$):
\begin{equation}
\label{eq:RelativisticAction}
S=
-mc\int_{\lambda_1}^{\lambda_2}d\lambda
\sqrt{g_{\alpha\beta}\bigl(x(\lambda)\bigr)
{\dot x}^\alpha(\lambda){\dot x}^\beta(\lambda)} \,,
\end{equation}
where, in local coordinates, $x^\alpha(\lambda)$ is the
worldline parametrised by $\lambda$ and $g_{\alpha\beta}$
are the metric components. Specialised to static metrics
\begin{equation}
\label{eq:StaticMetric}
g=f^2(\vec x)c^2dt^2-h_{ab}(\vec x)\,dx^adx^b \,,
\end{equation}
we have
\begin{equation}
\label{eq:RelativisticActionStatic}
S=
-mc^2\int_{t_1}^{t_2}dt
f\bigl(\vec x(t)\bigr)\sqrt{1-\frac{\hat h(\vec v,\vec v)}{c^2}}\,,
\end{equation}
where $\vec v:=(v^1,v^2,v^3)$ with $v^a=dx^a/dt$, and where
$\hat h$ is the ``optical metric'' of the space sections
$t=\mathrm{const}$:
\begin{equation}
\label{eq:OpticalMetric}
\hat h_{ab}:=\frac{h_{ab}}{f^2}\,.
\end{equation}
This is valid for all static metrics. Next we assume the
metric to be spatially conformally flat, i.e.,
$h_{ab}=h^2\,\delta_{ab}$, or equivalently
\begin{equation}
\label{eq:OpticalMetricConfFlat}
{\hat h}_{ab}={\hat h}^2\,\delta_{ab}\,,
\end{equation}
with $\hat h:=h/f$, so that the integrand (Lagrange function)
of (\ref{eq:RelativisticActionStatic}) takes the form
\begin{equation}
\label{eq:RelativisticLagrangianStaticConfFlat}
L=
-mc^2\,
f\bigl(\vec x(t)\bigr)\sqrt{1-{\hat h}^2\bigl(\vec x(t)\bigr)\frac{\vec v^2}{c^2}} \,,
\end{equation}
where ${\vec v}^2:=(v^1)^2+(v^2)^2+(v^3)^2$.

We note that spherically symmetric metrics are necessarily
spatially conformally flat (in any dimension and regardless
of whether Einstein's equations are imposed). In particular,
the Schwarzschild solution is of that form, as is manifest
if written down in isotropic coordinates:
\begin{equation}
\label{eq:SchwarzschildMetric-a}
g=\left[\frac{1-\frac{r_S}{r}}{1+\frac{r_S}{r}}\right]^2\,c^2dt^2
-\left[1+\frac{r_S}{r}\right]^4\bigl(dx^2+dy^2+dz^2\bigr)\,.
\end{equation}
Here $r_S$ is the Schwarzschild radius:
\begin{equation}
\label{eq:SchwarzschildMetric-b}
r_s:=GM/2c^2\,.
\end{equation}
Hence, in this case,
\begin{equation}
\label{eq:SchwarzschildMetric-c}
f=\frac{1-\frac{r_S}{r}}{1+\frac{r_S}{r}}\,,\quad
h=\left[1+\frac{r_S}{r}\right]^2\,,\quad
\hat h=\frac{\left[1+\frac{r_S}{r}\right]^3}{1-\frac{r_S}{r}}\,.
\end{equation}

Back to (\ref{eq:RelativisticActionStatic}), we now
approximate it to the case of weak gravitational fields
and slow particle velocities. For weak fields, Einstein's
equations yield to leading order:
\begin{equation}
\label{eq:LinearisedEinsteinSol}
f=1+\frac{\phi}{c^2}\,,\quad
h=1-\frac{\phi}{c^2}\,,\quad
\hat h=1-\frac{2\phi}{c^2}\,.
\end{equation}
Here $\phi$ is the Newtonian potential, i.e.
satisfies $\Delta\phi=4\pi GT_{00}/c^2$ where
$\Delta$ is the Laplacian for the flat spatial
metric. Inserting this in
(\ref{eq:RelativisticLagrangianStaticConfFlat})
the Lagrangian takes the leading-order form
\begin{equation}
\label{eq:LagrangianLeadingOrder}
L=
-mc^2+\tfrac{1}{2}mv^2-m\phi\,.
\end{equation}

We note that the additional constant $m^2c^2$ neither influences
the evolution of the classical nor of the quantum-mechanical
state. Classically this is obvious. Quantum-mechanically this
constant is inherited with opposite sign by the Hamiltonian:
\begin{equation}
\label{eq:Hamiltonian}
H=
mc^2+\tfrac{p^2}{2m}+m\phi\,.
\end{equation}
It is immediate that if $\psi(t)$ is a solution to the
time-dependent Schr\"odinger equation for this Hamiltonian,
then $\psi(t):=\exp\bigl(i\alpha(t)\bigr)\psi'(t)$
with $\alpha(t)=- t(mc^2/\hbar)$, where $\psi'$ solves the
time-dependent Schr\"odinger equation without the term
$mc^2$. But $\psi$ and $\psi'$ denote the \emph{same} time
sequence of states (rays). Hence we can just ignore this
term.\footnote{In \cite{Mueller.etal:2010} this term seems to
have been interpreted as if it corresponded to an inner
degree of freedom oscillating with Compton frequency,
therefore making up a ``Compton clock''. But as there is
no periodic change of state associated to this term, it
certainly does not correspond to anything like a clock
(whose state changes periodically) in this model.}

\subsection{Interferometry of freely falling atoms}
We now analyse the quantum mechanical coherences of the
centre-of-mass motion in a static and homogeneous gravitational
field, where we generalise to  $m_i\ne m_g$. Hence, instead of
(\ref{eq:LagrangianLeadingOrder}) (without the irrelevant
$mc^2$ term) we take
\begin{equation}
\label{eq:NR-Lagnrangian}
L=\tfrac{1}{2}m_i {\dot z}^2-m_ggz\,.
\end{equation}
Here we restricted attention to the vertical degree of
freedom, parametrised by $z$, where $\dot z:=dz/dt$.
This is allowed since the equation separates and implies
free evolution in the horizontal directions. The crucial
difference to  (\ref{eq:LagrangianLeadingOrder}) is that
we do not assume that $m_i=m_g$.

\begin{figure}[h]
\centering
\includegraphics[width=0.85\linewidth]{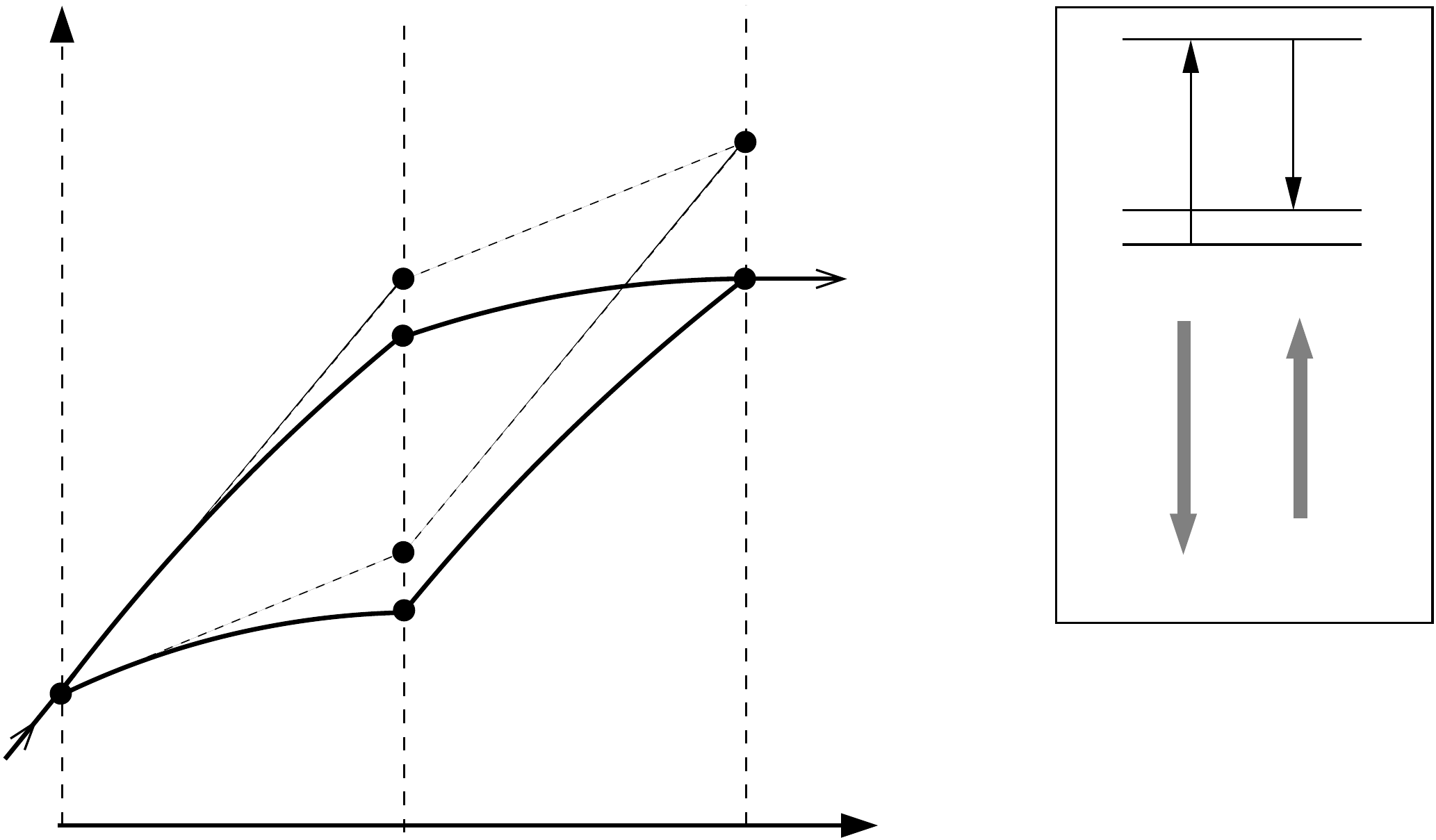}
\put(-102,0){$t$}
\put(-272,-10){$0$}
\put(-206,-10){$T$}
\put(-142,-10){$2T$}
\put(-272,167){$z$}
\put(-280,28){\footnotesize $A$}
\put(-200,90){\footnotesize $C$}
\put(-215,110){\footnotesize $C'$}
\put(-200,40){\footnotesize $B$}
\put(-215,60){\footnotesize $B'$}
\put(-134,113){\footnotesize $D$}
\put(-134,140){\footnotesize $D'$}
\put(-13,116){\tiny $\vert g_1\rangle$}
\put(-13,123){\tiny $\vert g_2\rangle$}
\put(-11,155){\footnotesize $\vert i\rangle$}
\put(-58,140){\footnotesize $\omega_1$}
\put(-26,140){\footnotesize $\omega_2$}
\put(-60,80){\footnotesize $\vec k_1$}
\put(-23,80){\footnotesize $\vec k_2$}
\put(-295,9){\tiny $\vert g_1\rangle$}
\put(-245,30){\tiny $\vert g_2\rangle$}
\put(-245,78){\tiny $\vert g_1\rangle$}
\put(-180,65){\tiny $\vert g_1\rangle$}
\put(-180,111){\tiny $\vert g_2\rangle$}
\put(-115,109){\tiny $\vert g_2\rangle$}
\caption{\label{fig:FigParallelogram}
Atomic interferometer with beam splitters at $A$ and $D$
and mirrors at $B$ and $C$, realised by $\pi/2$-- and $\pi$--pulses
of counter-propagating laser beams with $\vec k_1=-k_1\vec e_z$,
where $k_1=\omega_1/c$ and $\vec k_2=k_2\vec e_z$,
where $k_2=\omega_2/c$. $\vert g_{1{,}2}\rangle$ denote the
hyperfine doublet of ground states and $\vert i\rangle$ an
intermediate state via which the Raman transitions between
the ground states occur. The solid (bent) paths show the
classical trajectories in presence of a downward pointing
gravitational field $\vec g=-g\vec e_z$, the dashed (straight)
lines represent the classical trajectories for $g=0$.
The figure is an adaptation of Fig.\,9 in
\cite{Storey.CohenTannoudji:1994}.}
\end{figure}
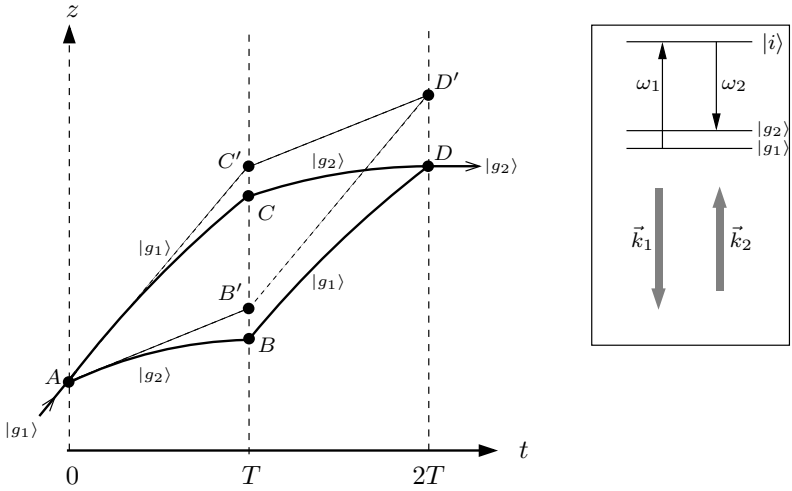
The situation described in \cite{Mueller.etal:2010}, which
is as in \cite{Storey.CohenTannoudji:1994}, is depicted in
Fig.\,\ref{fig:FigParallelogram}.
 A beam of Caesium atoms, initially in the lower state
$\vert g_1\rangle$ of the hyperfine doublet
$\vert g_1\rangle,\vert g_2\rangle$ of ground states, is
coherently split by two consecutive laser pulses which we may
take to act simultaneously at time $t=T$:
The first downward-pointing pulse with $\vec k_1=-k_1\vec e_z$, where
$k_1:=\omega_1/c$, elevates the atoms from the ground state
$\vert g_1\rangle$ to an intermediate state $\vert i\rangle$,
thereby transferring momentum $\Delta_{A1}\vec p=-\hbar k_1\vec e_z$.
The second upward-pointing pulse $\vec k_2=k_2\vec e_z$, where
$k_2:=\omega_2/c$, induces a transition from $\vert i\rangle$
to the upper level of the hyperfine-split ground state,
$\vert g_2\rangle$. The emitted photon is pointing
upwards so that the atom suffers a recoil by
$\Delta_{A2}\vec p=-\hbar k_2\vec e_z$. The total momentum
transfer at $A$ is the sum
$\Delta_A \vec p:=\Delta_{A1}\vec p+\Delta_{A2}\vec p=-\kappa\vec e_z$
with $\kappa=k_1+k_2=(\omega_1+\omega_2)/c$. At $A$ the pulses
are so adjusted that each atom has a 50\% chance to make this
transition
$\vert g_1\rangle\rightarrow\vert i\rangle\rightarrow\vert g_2\rangle$
and proceed on branch $AB$  and a 50\% chance to just stay in
$\vert g_1\rangle$ and proceed on branch $AC$. (So-called
$\pi/2$--pulse or atomic beam splitter.) At $B$ and $C$
the pulses are so adjusted that the atoms make transitions
$\vert g_2\rangle\rightarrow\vert i\rangle\rightarrow\vert g_1\rangle$
and $\vert g_1\rangle\rightarrow\vert i\rangle\rightarrow\vert g_2\rangle$
with momentum changes $\Delta_B\vec p=\kappa\vec e_z$ and
$\Delta_C\vec p=-\kappa\vec e_z$ respectively. Here the pulses
are so adjusted that these transitions occur almost with 100\%
chance. (So-called $\pi$--pulses or atomic mirrors.) Finally, at
$D$, the beam from $BD$, which is in state $\vert g_1\rangle$,
receives another $\pi/2$-pulse so that $50\%$ of it re-unites
coherently with the transmitted 50\% of the beam incoming from
$CD$ that is not affected by the  $\pi/2$--pulse at $D$.
In total, the momentum transfers on the upper and lower paths are,
respectively,
\begin{subequations}
\label{eq:MomentumTransfers}
\begin{alignat}{2}
&\label{eq:MomentumTransfers-upper}
\Delta_{\rm upper}\vec p&&=\Delta_C\vec p=
\underbrace{-\hbar\kappa\vec e_z}_{\text{at $C$}}\,,\\
\label{eq:MomentumTransfers-lower}
&\Delta_{\rm lower}\vec p&&=\Delta_A\vec p+\Delta_B\vec p+\Delta_D\vec p=
\underbrace{-\hbar\kappa\vec e_z}_{\text{at $A$}}
\underbrace{+\hbar\kappa\vec e_z}_{\text{at $B$}}
\underbrace{-\hbar\kappa\vec e_z}_{\text{at $D$}}\,.
\end{alignat}
\end{subequations}

Following \cite{Storey.CohenTannoudji:1994} we now wish to
show how to calculate the phase difference along the two
different paths using (\ref{eq:ExactPropagator}). For this
we need to know the classical trajectories.
\begin{rem}
\label{rem:GandHatG-1}
The classical trajectories are parabolic with downward
acceleration of modulus $\hat g$. If the trajectory is a
stationary point of the classical action we would have
$\hat g=g(m_g/m_i)$. However, the authors of
\cite{Mueller.etal:2010} contemplate the possibility
that violations of UGR could result in not making this
identification.%
\footnote{\label{footnote:GhatVersusGdash}
In \cite{Mueller.etal:2010} the quantity that we call
$\hat g$ is called $g'$.} Hence we proceed without specifying
$\hat g$ until the end, which also has the additional
advantage that we can at each stage cleanly distinguish
between contributions from the kinetic and contributions
from the potential part of the action. But we do keep in mind
that (\ref{eq:ExactPropagator}) only represents the right
dependence on $(z_a,z_b)$ if $z(t)=z_*(t)$ holds. This will
be a crucial point in our criticism to which we return later
on.
\end{rem}

Now, the unique parabolic orbit with downward acceleration $\hat g$
through initial event $(t_a,z_a)$ and final event $(t_b,z_b)$ is
\begin{equation}
\label{eq:OrbitGravField-1}
z(t)=z_a+v_a(t-t_a)-\tfrac{1}{2}\hat g(t-t_a)^2\,,
\end{equation}
where
\begin{equation}
\label{eq:OrbitGravField-2}
v_a:=\frac{z_b-z_a}{t_b-t_a}+\frac{1}{2}\hat g(t_b-t_a)\,.
\end{equation}
Evaluating the action along this path gives
\begin{equation}
\label{eq:ClassicalActionEvaluated}
\begin{split}
S_{\hat g}(z_b,t_b;z_a,t_a)
&=\frac{m_i}{2}\frac{(z_b-z_a)^2}{t_b-t_a}\\
&\mspace{20mu}-\frac{m_gg}{2}(z_b+z_a)(t_b-t_a)\\
&\mspace{20mu}+\frac{\hat g}{24}(t_b-t_a)^3(m_i\hat g-2m_gg)\,.\\
\end{split}
\end{equation}
\begin{rem}
\label{rem:GandHatG-2}
Note the different occurrences of $g$ and $\hat g$ in this equation.
We recall that $g$ is the parameter in the Lagrangian
(\ref{eq:NR-Lagnrangian}) that parametrises the gravitational
field strength, whereas $\hat g$ denotes the modulus of the
downward acceleration of the trajectory $z(t)$ along which
the atom actually moves, and along which the action is evaluated.
\end{rem}
\begin{rem}
\label{rem:KinericAndPotentialParts}
In (\ref{eq:ClassicalActionEvaluated}) it is easy to tell apart
those contributions originating from the kinetic term (first
term in (\ref{eq:NR-Lagnrangian}))
from those originating from the potential term (second
term in (\ref{eq:NR-Lagnrangian})): The former are
proportional to $m_i$, the latter to $m_g$.
\end{rem}
\begin{rem}
\label{rem:ChangeIntState}
Note also that on the upper path $ACD$ the atom changes the
internal state at $C$ and on the lower path $ABD$ it changes the
internal state at all three laser-interaction points $A$, $B$,
and $D$. According to Special Relativity, the atom also changes
its inertial mass at these points by an amount $\Delta E/mc^2$,
where $\Delta E\approx4\times 10^{-5}\,\mathrm{eV}$, which is a
fraction of $3\cdot 10^{-16}$ of Caesium's rest energy $mc^2$.%
\footnote{Recall that the ``second'' is defined to be the
duration of $\nu=9,192,631,770$ cycles of the hyperfine structure
transition frequency of Caesium-133. Hence, rounding up to three
decimal places,  the energy of this transition is
$\Delta E=h\times\nu = 4.136\cdot 10^{-15}\,\mathrm{eV\cdot s}\times
9,193,631,770\,\mathrm{s^{-1}}=3.802\cdot 10^{-5}\,\mathrm{eV}$.
The mass of Caesium is $m=132.905\,\mathrm{u}$, where
$1\,\mathrm{u}=931.494\cdot 10^{6}\,\mathrm{eV/c^2}$;
hence $mc^2=1.238\cdot 10^{11}\,\mathrm{eV}$ and
$\Delta E/mc^2=3.071\cdot 10^{-16}$.} Hence this change in
inertial mass, as well as a change in gravitational mass of
the same order of magnitude, can be safely neglected without
making any assumptions concerning the constancy of their
quotient $m_i/m_g$.
\end{rem}

Now we come back to the calculation of phase shifts.
According to (\ref{eq:ExactPropagator}), we need
to calculate the classical actions along the upper
and lower paths.
\begin{equation}
\label{eq:SumOfActions}
\begin{split}
\Delta S&=S(AC)+S(CD)-\bigl(S(AB)+S(BD)\bigr)\\
        &=\bigl(S(AC)-S(AB)\bigr)+\bigl(S(CD)-S(BD)\bigr)\,.\\
\end{split}
\end{equation}
\begin{rem}
Since the events $B$ and $C$ differ in time from $A$ by the
same amount $T$, and likewise $B$ and $C$ differ from
$D$ by the same amount $T$, we see that the last term on the
right-hand side of (\ref{eq:ClassicalActionEvaluated}) drops
out upon taking the differences in (\ref{eq:SumOfActions}),
since it only depends on the time differences but not on
the space coordinates. Therefore, the dependence on $\hat g$
also drops out of (\ref{eq:ClassicalActionEvaluated}), which
then only depends on $g$.
\end{rem}

The calculation of (\ref{eq:SumOfActions}) is now easy.
We write the coordinates of the four events $A,B,C,D$ in
Fig.\,\ref{fig:FigParallelogram} as
\begin{alignat}{2}
A&=(z_A,t_A=0)\,,\quad
B&&=(z_B,t_B=T)\,,\nonumber\\
\label{eq:LabelFourEvents}
C&=(z_C,t_C=T)\,,\quad
D&&=(z_D,t_D=2T)\,
\end{alignat}
and get
\begin{equation}
\label{eq:SumOfActionsEval}
\Delta S=
\frac{m_i}{T}(z_C-z_B)\bigl[z_B+z_C-z_A-z_D-g(m_g/m_i)T^2\bigr]\,.
\end{equation}
Since the curved (thick) lines in Fig.\,\ref{fig:FigParallelogram}
are the paths with downward acceleration of modulus $\hat g$,
whereas the straight (thin) lines correspond to the paths
without a gravitational field, the corresponding coordinates
are related as follows%
\footnote{\label{foot:StateChange}%
If we took into account the fact that the atoms change
their inner state and consequently their inertial mass $m_i$, we
should also account for the possibility that the quotient $m_i/m_g$
may change. As a result, the magnitude of the downward
acceleration $\hat g$ may depend on the inner state. These
quantities would then be labelled by indices $1$ or $2$,
according to whether the atom is in state $\vert g_1\rangle$ or
$\vert g_2\rangle$, respectively. The modulus of the downward acceleration along
$AC$ and $BD$ is then ${\hat g}_1$, and ${\hat g}_2$ along $AB$ and
$CD$. Also, the momentum transfers through laser interactions are
clearly as before, but if converted into velocity changes by using
momentum conservation one has to take into account that the inertial
mass changes during the interaction. However, the $z$-component of
the velocity changes at $A$ (for that part of the incoming beam at
$A$ that proceeds on $AB'$) and $C'$ will still be equal in magnitude
and oppositely directed to that at $B'$, as one can easily convince
oneself; its magnitude being
$\Delta v=\big(1-\frac{m_{i1}}{m_{i2}}\big)v_A+\hbar\kappa/m_{i2}$,
where $v_A$ is the incoming velocity in $A$. As a result,
it is still true that for laser-induced Raman transitions with
momentum transfer $\hbar\kappa$ at $t=0$ and $t=T$ the two beams from
$C'$ and $B'$ meet at time $t=2T$ at a common point,
which is $z_{D'}=z_A+2v_AT-\frac{\hbar\kappa}{m_{i2}}+\big(\frac{m_{i1}}{m_{i2}}-1\big)v_AT$.
Switching on the gravitational field has the effect that
$z_C=z_{C'}-\frac{1}{2}{\hat g}_1T^2$ and $z_B=z_{B'}-\frac{1}{2}{\hat g}_2T^2$,
but that the beam from $C$ arrives after time $T$ at
$z^{(C)}_{D}=z_{D'}-\frac{1}{2}({\hat g}_1+{\hat g}_2)T^2-
\frac{m_{i1}}{m_{i2}}{\hat g}_1T^2$,
whereas the beam from $B$ arrives after time $T$ at
$z^{(B)}_{D}=z_{D'}-\frac{1}{2}({\hat g}_1+{\hat g}_2)T^2-
\frac{m_{i2}}{m_{i1}}{\hat g}_2T^2$,
which differs from the former by $\Delta z_D:=z^{(C)}_{D}-z^{(B)}_{D}=
T^2\Big(\frac{m_{i2}}{m_{i1}}{\hat g}_2-\frac{m_{i1}}{m_{i2}}{\hat g}_1\Big)$.
Assuming that ${\hat g}_1=(m_{g1}/m_{i1})g$ and  ${\hat g}_2=(m_{g2}/m_{i2})g$,
this is  $\Delta z_D=(m_{i2}m_{g2}-m_{i1}m_{g1})gT^2/(m_{i1}m_{i2})$
which, interestingly, vanishes iff the \emph{product}
(rather than the quotient) of inertial and gravitational mass stays
constant. If the quotient stays approximately constant and so that
${\hat g}_1={\hat g}_2=:\hat g$, we write $m_{i2}/m_{i1}=1+\varepsilon$,
with $\varepsilon=\Delta E/m_{i1}c^2$, and get to first order in
$\varepsilon$ that $\Delta z_D\approx 2\varepsilon\hat gT^2$.}:
\begin{equation}
\label{eq:LabelFourEventsPrimed}
\begin{split}
A'&=\bigl(z_{A'}=z_A, t_{A'}=0\bigr)\,,\\
B'&=\bigl(z_{B'}=z_B+\tfrac{1}{2}{\hat g}T^2\,,\,t_{B'}=T\bigr)\,,\\
C'&=\bigl(z_{C'}=z_C+\tfrac{1}{2}{\hat g}T^2\,,\,t_{C'}=T\bigr)\,,\\
D'&=\bigl(z_{D'}=z_D+2{\hat g}T^2           \,,\,t_{D'}=2T\bigr)\,.\\
\end{split}
\end{equation}
Hence
\begin{equation}
\label{eq:FourEventRelation}
z_{B}+z_{C}-z_{A}-z_{D}
=z_{B'}+z_{C'}-z_{A'}-z_{D'}+\hat gT^2
=\hat gT^2\,,
\end{equation}
where $z_{B'}+z_{C'}-z_{A'}-z_{D'}=0$ simply follows from the fact
that $A'B'C'D'$ is a parallelogram. Moreover, for the difference
$(z_C-z_B)$ we have
\begin{equation}
\label{eq:TwoEventRelation}
z_{C}-z_{B}=z_{C'}-z_{B'}=\frac{\hbar\kappa}{m_i}T\,,
\end{equation}
where the last equality follows from the fact that
along the path $AB'$ the atoms have an additional
momentum of $-\hbar\kappa{\vec e}_z$ as compared to
the atoms along the path $AC'$; compare
(\ref{eq:MomentumTransfers-lower}). Using
(\ref{eq:FourEventRelation}) and (\ref{eq:TwoEventRelation})
in (\ref{eq:SumOfActionsEval}), we get:
\begin{equation}
\label{eq:SumOfActionsEvalFinal}
\Delta S=
\hbar\kappa T^2\,\bigl[\hat g-g(m_g/m_i)\bigr]\,.
\end{equation}
As advertised in Remarks~\ref{rem:GandHatG-1},
\ref{rem:GandHatG-2}, and \ref{rem:KinericAndPotentialParts},
we can now state individually the contributions to the
phase shifts from the kinetic and the potential parts:
\begin{subequations}
\label{eq:PhaseShifts}
\begin{alignat}{2}
\label{eq:PhaseShift-Time}
&(\Delta\phi)_{\rm time}&&\,=\,+\,\kappa T^2\,\hat g\,,\\
\label{eq:PhaseShift-Redshift}
&(\Delta\phi)_{\rm redshift}&&\,=\,-\,\kappa T^2\,g(m_g/m_i)\,.
\end{alignat}
Here ``time'' and ``redshift'' remind us that, as explained in
Section\,~\ref{sec:GR-Story}, the kinetic and potential energy
terms correspond to the leading-order Special-Relativistic time
dilation (Minkowski geometry) and the influence of gravitational
fields, respectively.

Finally we calculate the phase shift due to the laser
interactions at $A,B,C$ and $D$. For the centre-of-mass
wave function to which we restrict attention here, only
the total momentum transfers matter which were already
stated in (\ref{eq:MomentumTransfers}). Hence we get
for the phase accumulated on the upper path $ACD$ minus
that on the lower path $ABD$:
\begin{alignat}{1}
(\Delta\phi)_{\rm light}
&\,=\,\frac{1}{\hbar}\bigl\{
(\Delta_C\vec p)\cdot\vec z_C-
(\Delta_A\vec p)\cdot\vec z_A-
(\Delta_B\vec p)\cdot\vec z_B-
(\Delta_D\vec p)\cdot\vec z_D\bigr\}
\nonumber \\
\label{eq:PhaseShift-Light}
&\,=\,-\kappa (z_B+z_C-z_A-z_D)=-\,\kappa\hat gT^2 \,,
\end{alignat}
\end{subequations}
where we used (\ref{eq:FourEventRelation}) in the last step.

Taking the sum of all three contributions in
(\ref{eq:PhaseShifts}) we finally get
\begin{equation}
\label{eq:PhaseShift-Total}
\Delta\phi=-\,\frac{m_g}{m_i}\cdot g\cdot\kappa\cdot T^2\,.
\end{equation}
This is fully consistent with the more general formula
derived by other methods (no path integrals) in
\cite{Laemmerzahl:1996}, which also takes into account
possible inhomogeneities of the gravitational field.

\subsection{Atom interferometers testing UFF}
\label{sec:AtomInterferometersUFF}
Equation (\ref{eq:PhaseShift-Total}) is the main result of
the previous section. It may be used in various ways. For
given knowledge of $(m_g/m_i)$ a measurement of $\Delta\phi$
may be taken as a measurement of $g$. Hence the atom
interferometer can be used as a gravimeter. However, in the
experiments referred to in \cite{Mueller.etal:2010} there
was another macroscopic gravimeter nearby consisting of a
freely falling corner-cube retroreflector monitored by a laser
interferometer. If $M_i$ and
$M_g$ denote the inertial and gravitational mass of the
corner cube, its acceleration in the gravitational field
will be $\tilde g=(M_g/M_i)g$. The corner-cube accelerometer
allows to determine this acceleration up to
$\Delta\tilde g/\tilde g<10^{-9}$~\cite{Peters.Chung.Chu:2001}.
Hence we can write
(\ref{eq:PhaseShift-Total}) as
\begin{equation}
\label{eq:PhaseShift-TotalAlt}
\Delta\phi=-\,\frac{m_g}{m_i}\cdot\frac{M_i}{M_g}\cdot
\bigl(\tilde g\kappa T^2\bigr)\,,
\end{equation}
in which the left-hand side and the bracketed terms
on the right-hand side are either known or measured.
Using the E\"otv\"os ratio (\ref{eq:DefEoetvoesFactor})
for the Caesium atom ($A$) and the reference cube ($B$)
\begin{equation}
\label{eq:EoetvoesRatio}
\eta(\text{atom},\text{cube})
=2\cdot\frac{(m_g/m_i)-(M_g/M_i)}{(m_g/m_i)+(M_g/M_i)} \,,
\end{equation}
we have
\begin{equation}
\label{eq:MassQuotient}
\frac{m_gM_i}{m_iM_g}=\frac{2+\eta}{2-\eta}= 1+\eta +\mathcal{O}(\eta^2)\,.
\end{equation}
Hence, to first order in $\eta:=\eta(\text{atom},\text{cube})$,
we can rewrite~(\ref{eq:PhaseShift-TotalAlt}):
\begin{equation}
\label{eq:PhaseShift-TotalEta}
\Delta\phi=-\,(1+\eta)\cdot
\bigl(\tilde g\kappa T^2\bigr)\,.
\end{equation}
This formula clearly shows that measurements of phase shifts
can put upper bounds on $\eta$ and hence on possible
violations of UFF.

The experiments \cite{Peters.Chung.Chu:1999,Peters.Chung.Chu:2001}
reported in \cite{Mueller.etal:2010} led to a measured redshift per
unit length (height) which, compared to the predicted values, reads
as follows:
\begin{subequations}
\label{eq:RedShiftPerUnitHeight}
\begin{alignat}{3}
\label{eq:RedShiftPerUnitHeight-meas}
&\zeta_{\rm meas}
&&:\,=\,\frac{-\Delta\phi}{\kappa T^2c^2}
&&\,=\,(1.090\,322\,683\pm 0.000\,000\,003)\times 10^{-16}\cdot \mathrm{m}^{-1}\,,\\
&\zeta_{\rm pred}
&&:\,=\,\,\tilde g/c^2
&&\,=\,(1.090\,322\,675\pm 0.000\,000\,006)\times 10^{-16}\cdot \mathrm{m}^{-1}\,.
\end{alignat}
\end{subequations}
This implies an upper bound of
\begin{equation}
\label{eq:EoetvoesTests-MuellerEtAl}
\eta(\text{atom},\text{cube})=\frac{\zeta_{\rm meas}}{\zeta_{\rm pred}}-1<(7\pm 7)\times 10^{-9}\,,
\end{equation}
which is more than four orders of magnitude worse (higher)
than the lower bounds obtained by more conventional methods
(compare \eqref{eq:EoetvoesTests}). However, it should be
stressed that here a comparison is made between a macroscopic
body (cube) and a genuine quantum-mechanical system (atom)
in a superposition of centre-of-mass eigenstates, whereas
other tests of UFF use macroscopic bodies describable by
classical (non-quantum) laws.

\subsection{Atom interferometers testing URS?}
\label{sec:AtomInterferometersUCR}
The foregoing interpretation seems straightforward and is
presumably uncontroversial; but it is not the one adopted by the
authors of \cite{Mueller.etal:2010}. Rather, they claim that a
measurement of $\Delta\phi$ can, in fact, be turned into
an upper bound on the parameter $\alpha_{\rm RS}$ which,
according to them, enters the formula (\ref{eq:PhaseShift-TotalEta})
for the predicted value of $\Delta\phi$ just in the same fashion
as does $\eta$. Then, since other experiments constrain $\eta$ to
be much below the $10^{-9}$ level, the very same reasoning as
above now leads to the upper bound (\ref{eq:EoetvoesTests-MuellerEtAl})
for $\alpha_{\rm RS}$ rather than $\eta$, which now implies a dramatic
improvement of the upper bound (\ref{eq:UpperBoundAlphaRS})
by four orders of magnitude!

However, the reasoning given in
\cite{Mueller.etal:2010,Mueller.etal:2010b} for how $\alpha_{\rm RS}$
gets into (\ref{eq:PhaseShift-TotalEta}) seems theoretically
inconsistent. It seems to rest on the observation that
(\ref{eq:PhaseShift-Time}) cancels with (\ref{eq:PhaseShift-Light})
\emph{irrespectively of whether $\hat g=(m_g/m_i)g$ or not}, so that
\begin{equation}
\label{eq:PhaseDueToRedshift}
\Delta\phi=(\Delta\phi)_{\rm redshift}\,.
\end{equation}
Then they simply assumed that if violations of UGR
existed $(\Delta\phi)_{\rm redshift}$, and hence
$\Delta\phi$, simply had to be multiplied by $(1+\alpha_{\rm RS})$.
(Our $\alpha_{\rm RS}$ is called $\beta$ in \cite{Mueller.etal:2010}.)

\begin{rem}
The cancellation of \eqref{eq:PhaseShift-Time} with
\eqref{eq:PhaseShift-Light} for $\hat g\ne (m_g/m_i)g$ is
formally correct but misleading.  The reason is apparent from
(\ref{eq:ExactPropagator}): The action has to be evaluated along
the solution $z_*(t)$ in order to yield the dynamical phase of
the wave function. Evaluating it along any other trajectory will
not solve the Schr\"odinger equation. Therefore, whenever
$\hat g\ne (m_g/m_i)g$, the formal manipulations performed
are physically, at best, undefined. On the other hand, if
$\hat g=(m_g/m_i)g$, then according to  (\ref{eq:PhaseShifts})
\begin{equation}
\label{eq:TrippleCancellations}
(\Delta\phi)_{\rm time}=-
(\Delta\phi)_{\rm redshift}=-
(\Delta\phi)_{\rm light} \,,
\end{equation}
so that we may just as well say that the total phase is entirely
due to the interaction with the laser, i.e. that we have instead
of (\ref{eq:PhaseDueToRedshift})
\begin{equation}
\label{eq:PhaseDueToLight}
\Delta\phi=(\Delta\phi)_{\rm light}
\end{equation}
and no $\alpha_{\rm RS}$ will enter the formula for the phase.
\end{rem}
\begin{rem}
The discussion in Section\,\ref{sec:EnergyCons-UFF_UGR} suggests
that violations of UGR are quantitatively constrained by
violations of UFF if energy conservation holds. If the upper
bounds for violations of UFF are assumed to be on the $10^{-13}$
level (compare~(\ref{eq:EoetvoesTests})), this means that violations
of UGR cannot exceed the $10^{-9}$ level for magnetic interactions.
Since the latter is just the new level allegedly reached by the argument
in \cite{Mueller.etal:2010}, we must conclude by Nordtvedt's gedanken
experiment that the violations of UGR that are effectively excluded by
the argument of \cite{Mueller.etal:2010} are those also violating
energy conservation.
\end{rem}

\begin{rem}
Finally we comment on the point repeatedly stressed in
\cite{Mueller.etal:2010} that (\ref{eq:ExactPropagator}) together
with the relativistic form of the action (\ref{eq:RelativisticAction})
shows that the phase change due to the free dynamics
$(\Delta\phi)_{\rm free}:=\Delta\phi-(\Delta\phi)_{\rm light}$,
which in the leading order approximation is just the $dt$-integral
over (\ref{eq:LagrangianLeadingOrder}), can be written as the
integral of the eigentime times the constant Compton frequency
$\omega_C=m_ic^2/\hbar$:
\begin{equation}
\label{eq:ComptonIntegral}
(\Delta\phi)_{\rm free}=\omega_C\int d\tau \,.
\end{equation}
The authors of \cite{Mueller.etal:2010} interpret this
as timing the length of a worldline by a ``Compton clock''
(\cite{Mueller.etal:2010}, p.\,927).
\footnote{``The essential realisation of this Letter is that the
non-relativistic formalism hides the true quantum oscillation
frequency $\omega_C$.'' (\cite{Mueller.etal:2010}, pp.\,928 and 930)}
For Caesium atoms this frequency is about
$2\times 10^{26}\cdot\mathrm{s}^{-1}$, an enormous value. However,
there is no periodic change of any physical state associated to this
frequency, unlike in atomic clocks, where the beat frequency of two
stationary states gives the frequency by which the superposition
state (ray in Hilbert space) periodically recurs. Moreover, the
frequency $\omega_C$ apparently plays no r\^ole in any of the
calculations performed in \cite{Mueller.etal:2010}, nor is it
necessary to express $\Delta\phi$ in terms of known quantities.
The mere fact that the phase difference of two hypothetical 
clocks ticking at Compton frequency would just be the one 
discussed above does clearly not imply that in the situation 
at hand we \emph{are} dealing with genuine clocks. Hence the 
reference to a ``Compton clock'' is unwarranted and likewise 
the stipulation that if violations of UGR existed they would 
register in the phase $\Delta\phi$ by multiplying it with 
$(1+\alpha_{\rm RS})$. 
\end{rem}

\section{Conclusion}
I conclude from the discussion of the previous section that the
arguments presented in \cite{Mueller.etal:2010} are not sufficient 
to show a direct proportionality of $\Delta\phi$ with 
$(1+\alpha_{\rm RS})$, which is the basis of its main argument. 
As a result, \cite{Mueller.etal:2010}
does not provide sufficient reason to claim an improvement on 
upper bounds on possible violations of UGR---and hence on all 
of EEP---by four orders of magnitude. This would indeed have been 
a major achievement, as UCR/UGR is by far the least well-tested part
of EEP, which, to stress it once more, is the connector between
gravity and geometry. Genuine quantum tests of EEP are most welcome
and the experiment described in \cite{Mueller.etal:2010} is certainly
a test of UFF, but not of UGR. As a test of UFF it is still more than
four orders of magnitude away from the best non-quantum torsion-balance
experiments. However, one should stress immediately that it puts bounds
on the E\"otv\"os factor relating a classically describable piece of
matter to an atom in a superposition of spatially localised states, and
as such it remains certainly useful. On the other hand, as indicated
in Fig.\,\ref{fig:FigParallelogram}, the atoms are in energy eigenstates
$\vert g_{1,2}\rangle$ between each two interaction points on the upper
and on the lower path. Hence we do \emph{not} have a genuine quantum
test of UGR where a quantum clock (being in a superposition of energy
eigenstates) is coherently moving on two different worldlines.
This seems to have been the idea of \cite{Mueller.etal:2010} when
calling each massive system a ``Compton clock''. It remains to be
seen whether and how this idea can eventually be realised with real
physical quantum clocks, i.e.  quantum systems whose state (ray!) is
periodically changing in time. This has also recently been suggested 
in \cite{Samuel.Sinha:2011} and \cite{Zych.etal:2011}.

\bibliographystyle{plain}
\bibliography{RELATIVITY,QM} 
\end{document}